\documentclass[11pt]{article}
\usepackage[utf8]{inputenc}
\usepackage{authblk}
\usepackage{amsmath}
\usepackage[pagebackref=true]{hyperref} 
\usepackage{graphicx}
 \renewcommand*\backref[1]{\ifx#1\relax \else (Cited on #1) \fi}
\newcommand{\ITEM}{restaurant }
\newcommand{\ITEMS}{restaurants }

\title{Inverse Propensity Score based offline estimator for deterministic ranking lists using position bias}
\author[1]{Nick Wood}
\author[1]{Sumit Sidana}
\affil[1]{Just Eat Takeaway.com}
\date{\today}

\begin{document}

\maketitle

\section{Introduction}

The mission of Just Eat Takeaway is to empower users’ every food moment. A big part of fulfilling that mission is ensuring that we show the right restaurants to the right users. To do this, we design large-scale machine learning recommendation systems that can learn which types of users tend to enjoy which types of restaurants.

A crucial part of this process is being able to compare the quality of different recommendation systems, in order to decide which is most effective. The gold standard approach to this evaluation problem is A/B testing - allow the different systems to recommend items to randomly selected groups of users, and compare their relative performance.

However, A/B testing is both slow, taking time to reach sufficient statistical power, and expensive \cite{Gilotte_2018, Agarwal_2017}  if we serve poor recommendations to a subset of users.

Therefore, evaluating recommender systems in offline fashion without running A/B tests is of significant importance. These offline approaches avoid serving experimental recommenders to users, and instead evaluates them using old online data, where users were served recommendations from a different system. Metrics such as Root Mean Squared Error \cite{bennett2007netflix} Mean Average Precision, Normalized Discounted Cumulative Gain, Mean Reciprocal Rank and Hit Rate have been used repeatedly for offline evaluation.

All of these evaluation approaches are prone to selection, presentation and position biases \cite{rosetti_recsys_16, olivier_19} and therefore the results from offline evaluation are often inaccurate, meaning the evaluated algorithms perform differently when served online than the offline evaluation predicted.

 For this reason, “offline” approaches to evaluation have been subject to intense research \cite{phdthesis_olivier} with the goal of coming up with offline performance indicators, which match the online performance.  Counterfactual off-policy evaluation ($OPE$, henceforth) based approaches often make use of Inverse Propensity Score, also known as importance sampling \cite{mcbook}. However, IPS requires us to use a stochastic logging policy. In practice, this means that our recommender systems output a probability distribution over all the possible items that could be recommended, and we choose our recommendations by drawing them from this distribution. Our recommender will assign high probabilities to items it thinks are good recommendations, and low probabilities to poor items, but we will occasionally serve very different recommendations to those the recommender thinks is best.
 
 If we want to evaluate a different recommender using the data acquired, this randomness is crucial, as it means we will sometimes have recommendations in our logged data that are close to what our new recommender would have selected.

 However, introducing this randomness comes at a clear monetary cost, and it can be hard to justify paying this concrete cost for the less well defined benefit of faster recommender iteration. Ideally, we would like some way to evaluate non-random (deterministic) recommender systems offline. In the next section, we describe our approach to the counterfactual offline estimator we used for measuring the performance of deterministic ranking lists by making use of position-bias model \cite{10.1145/3159652.3159732}.

\section{Offline estimation using a Position-Bias Model}
The propensity used in IPS is the probability that an item was selected to be displayed to a user. Position-bias modelling, in particular the examination model, gives us an alternative definition of item propensity. The examination model imagines that when a user sees a list of recommendations, they randomly choose a set of items to examine.

Each list position, k,  has an associated examination probability, $P_k(E)$, which usually is at its highest at the top of the list, and then decays to 0 for 
items in lower positions. From the list of examined items, the user chooses items which are relevant to them and clicks those which are. This process is modelled by assigning each item-user pair a probability of relevance $P_{i,u}(r)$.

We can use expectation maximisation \cite{Dempster} on logged clicked data to measure $P_k(E)$, the probability that an item is examined in each position \cite{10.1145/3159652.3159732}. Despite its simplicity, the examination model has been shown to predict user behaviour well \cite{2015Chuklin}, and we have used it with success on our ranking data in other applications. 

$P_k(E)$ gives us the probability distribution of restaurants being examined from top to bottom. We replace the propensity of an item being shown to a user with the propensity that they examine it. We argue that even if the ranked list is deterministic in nature, the user actually examines/sees the restaurant as a probability and under this assumption, the traditional IPS estimator still applies. So, we are doing IPS, but we’re doing it with estimated propensities, which are coming from an examination model \cite{10.1145/3159652.3159732}.

For the new policy, we get scores from new algorithm, rank \ITEMS according to those scores and find what position $p$ the \ITEM holds in this ranking according to the new policy and then find the position-bias for each \ITEM in this new ranking. For the logging policy, we find the position $k$ of each \ITEM in which it was originally shown and then find its position-bias. Hence, our estimator becomes as follows:

\begin{equation}
\begin{gathered}
        V_{ips}(\pi_e; D_0) = \frac{1}{n} \sum_{i = 1}^{n}\sum_{k=1}^{K}\frac{\pi_e(a_{i}|x_i)}{\pi_0(a_{i}|x_i)} \alpha_k r_{i,k}\\
        \text{where,}\\
        n=\text{number of all restaurants spread over all sessions}\\
        K = \text{maximum position on which click happened in that session}\\
        \pi_e(a_{i}|x_i) = \text{propensity of treatment policy for \ITEM $a_i$ for given user $x_i$} \\
        \pi_0(a_{i}|x_i) = \text{propensity of control policy for \ITEM $a_i$ for given user $x_i$}\\
        \alpha_k r_{i,k} = \text{the reward of item i at position k (dcg, number of clicks,  precision)}
\end{gathered}
\end{equation}

We also assume that the users come from the top and actions below the last order/click can be ignored. We only consider those restaurants above the clicked restaurant as true negative samples, as they were likely to have been seen. This assumption is similar to that of the RIPS estimator 
\cite{McInerney_2020}. Hence our estimator further becomes:

\begin{equation}
\begin{gathered}
\label{eq: pb_ope}
    V_{ips}(\pi_e; D_0) = \frac{1}{n} \sum_{i = 1}^{n}\sum_{k=1}^{K}\frac{\theta_p}{\theta_k} \alpha_k r_{i,k}\\
    \text{where}, \\
    \theta_p = \text{position-bias-at-position-p}\\
    \theta_k = \text{position-bias-at-position-k}\\
    \end{gathered}
\end{equation}

 $K = max(k)$ is the maximum position at which the click/order happened in each session and $n$ is the total number of sessions, in which at least one \ITEM was shown in the ranked list.

There is also a problem associated with equation \ref{eq: pb_ope} that most of the restaurants at lower ranks have a probability close to 0 of being examined. \cite{Sachdeva_2020} and \cite{SwaminathanKADL16} address this sort of problem. This issue will increase the variance of our estimator, but as our position bias decays only slowly to 0, we have not found significant variance problems.

\section{Experiments}

In this section, we describe the experiments done using two different A/B tests. We first compute our OPE score given by equation \ref{eq: pb_ope} for the novel recommender in both the experiments and compare it with the actual CTR obtained for the variant recommender.

\subsection{Multi-Objective Optimization Ranking Experiments}
In this section, we evaluate a deterministic multi-objective ranking recommender for ranking restaurants. We ran an A/B test comparing a control 2-objective control recommender which combined an objective function for new restaurants as well as a rule based objective focussing on distance and restaurant quality, which resulted in a blended ranking between these two objectives. The variant recommender combines 3 objectives: new \ITEMS, rule-based, and an additional personalization objective.

To test the quality of our OPE, we take the ranking sessions and clicks from the control experimental data and compute the counterfactual ranking that would have been served by the variant policy in this session. We can then use our estimator and the logged feedback to evaluate the offline quality of the new policy. By comparing this measurement to the real variant policy performance in the A/B test, we can test the quality of our OPE framework.

\begin{figure*}[!ht]
  \begin{center}
      \includegraphics[width=\textwidth]{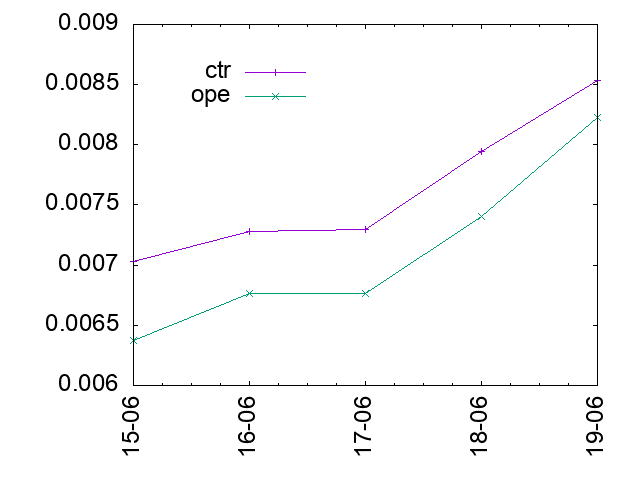}
      \vspace{-4mm}
      \caption{ope vs ctr for 11 days of A/B test of multi-objective optimization based ranking }
    \label{fig:ctr_ope_correlation_moon}
  \end{center}
\end{figure*}

\subsubsection{Independent Assumption}
In this experiment, we use the independence assumption. This also helps in making the action space tractable. Every \ITEM click is taken as independent from other \ITEM clicks. This means that one single appearance of a \ITEM in the ranked list is one data point in the experiment. The number of all such appearances is denoted by n in equation \ref{eq: pb_ope} Then, CTR definition is modified to:

\begin{equation}
\begin{gathered}
        CTR = \frac{1}{n} \sum_{i = 1}^{n}c\\
        \text{where,}\\
        c = \text{number of clicks in that session}
\end{gathered}
\end{equation}

Day by day results shown in Figure \ref{fig:ctr_ope_correlation_moon} show a very strong correlation between the $OPE$ estimated from control logs and $CTR$ value computed from treatment logs. There is a constant difference between the $CTR$ and $OPE$ value, which we are still trying to explain.

\subsection{Text Search Ranking Experiments}

\begin{figure*}[!ht]
  \begin{center}
      \includegraphics[width=\textwidth]{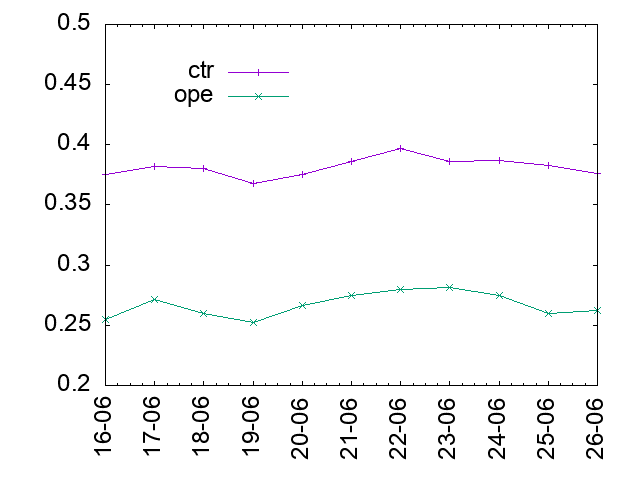}
      \vspace{-4mm}
      \caption{ope vs ctr for 11 days of A/B test of text search ranking }
    \label{fig:ctr_ope_correlation_text_search}
  \end{center}
\end{figure*}

In text search, users come in and input a text query in the text box and the ranked list of \ITEMS are shown. The $K = max(k)$ in equation \ref{eq: pb_ope} is the maximum position at which the click/order happened in each session and $n$ is the total number of sessions in which search was done and there was at least one result returned. We first run the position bias examination model on 1 month of logs in order to obtain $P_k(E)$.

Then we take logs from control and treatment and join these sessions on the text query and \ITEM to compute the position of each restaurant which was clicked in control. For control, we already have the position, but, for treatment, we take the median of all the possible positions,  which we end up as a result of the join. 
Then, we compare it with actual click-through-rate of each day of the treatment. For computing the OPE, we use equation \ref{eq: pb_ope}

Actual Click-Through-Rate is given by:

\begin{equation}
\begin{gathered}
    CTR =  \sum_{i = 1}^{n}\sum_{k=1}^{K}\frac{c}{v}    \\
    \text{where,}\\
    c = \text{$\#$clicks in that session}\\
    v = \text{$\#$views in that session}
\end{gathered}
\end{equation}

$K$ as before is the maximum position at which the click happened in that session. By using the maximum position, we make the clicks in ranked list a bit more dependent on each other. This kind of assumption is also similar to \cite{li_shuai}.

The results are shown in Figure \ref{fig:ctr_ope_correlation_text_search}, which shows day by day comparison of $CTR$ and $OPE$. As evident from the Figure, $OPE$ is strongly correlated with the actual click-through-rate obtained in treatment. There is a constant difference between the $OPE$ and actual $CTR$, which means $OPE$ is a bit underestimated compared to the actual $CTR$. Variance of OPE is also less compared to typical IPS estimator. This could have been due to the position-bias values not converging properly. This could also have been due to the fact that control and treatment policies don't differ a lot from each other. At this time, we are still investigating this further.

\section{Conclusion}
In this work, we presented a novel way of computing IPS using a position-bias model for deterministic logging policies. This technique significantly widens the policies on which OPE can be used. We validated this technique using two different experiments on industry-scale data. The OPE results are clearly strongly correlated with the online results, with some constant bias. The estimator requires the examination model to be a reasonably accurate approximation of real user behaviour.

\section{Acknowledgments}
We would like to thank Lauren Rodney, Erdem Biyik, Hakan Simsek, Liza Chukreeva and Rishi Kumar for helping us run these experiments. We would also like to thank Sabrican Özan, Max Knobbout and Boris Slesar for helping us review this article. Finally, we would like to thank Ozlem Ozer for sponsoring this project.

\bibliographystyle{plain}
\bibliography{counterfactual_off_policy_evaluation}

\end{document}